\documentclass[aps,article,prl,preprintnumbers,amsmath,amssymb,twocolumn,superscriptaddress,showpacs,floatfix]{revtex4}

\usepackage{epsfig}
\usepackage{dcolumn}
\usepackage{bm}
\usepackage{color}

\begin{document}
\title{Universal Flow-Driven Conical Emission in 
Ultrarelativistic Heavy-Ion Collisions}

\author{Barbara Betz}
\affiliation{Department of Physics, Columbia University, 
New York, 10027, USA}
\author{Jorge Noronha}
\affiliation{Department of Physics, Columbia University, 
New York, 10027, USA}
\author{Giorgio Torrieri}
\affiliation{Department of Physics, Columbia University, 
New York, 10027, USA}
\affiliation{Frankfurt Institute for Advanced Studies (FIAS), 
Frankfurt am Main, Germany}
\author{Miklos Gyulassy}
\affiliation{Department of Physics, Columbia University, 
New York, 10027, USA}
\author{Dirk H.\ Rischke}
\affiliation{Frankfurt Institute for Advanced Studies (FIAS), 
Frankfurt am Main, Germany}
\affiliation{Institut f\"ur Theoretische Physik, 
Johann Wolfgang Goethe-Universit\"at, Frankfurt am Main, Germany}

\begin{abstract}

The double-peak structure observed in soft-hard hadron correlations is
commonly interpreted as a signature for a Mach cone generated by a
supersonic jet interacting with the hot and dense medium created 
in ultrarelativistic heavy-ion collisions.
We show that it can also arise due to averaging over many jet events
in a transversally expanding background. We find that the jet-induced away-side yield does not depend on the details of the 
energy-momentum deposition in the plasma, the jet velocity, or the system size. Our claim can be experimentally tested by
comparing soft-hard correlations induced by heavy-flavor jets 
with those generated by light-flavor jets.

\end{abstract}

\maketitle


RHIC data have convincingly shown \cite{whitebrahms,whitephenix,whitephobos,whitestar} that a hot and
dense medium is created in ultrarelativistic collisions of heavy ions.
This medium is most likely the so-called quark-gluon plasma (QGP),
predicted by quantum chromodynamics to be the hot and dense phase 
of strongly interacting matter \cite{QGP}. 
The QGP was found to behave like an almost perfect fluid 
and to be opaque to jets created in the initial
stage of the collision \cite{Gyulassy:2004zy}.  
This raises the possibility of using the energy deposited by the jet in the medium, 
observable by correlations of soft and hard particles, as a probe of
the medium's properties. 

The experimental soft-hard correlation function 
\cite{Adams:2005ph,Adler:2005ee,2pcPHENIX,UleryPRL} 
exhibits an interesting double-peak structure at angles opposite to the
trigger jet. It has been suggested
\cite{Stoecker:2004qu,CasalderreySolana:2004qm} 
that such a structure is evidence for Mach cones.
If the QGP is an opaque low-viscosity fluid, Mach cones
result from the interference of sound waves generated
by the energy deposited by a supersonic jet \cite{Landau}.

The Mach cone leads to an excess of low-$p_T$ hadrons being 
emitted at an angle that is roughly $\pi - \phi_M$ with respect to the
trigger jet, where the Mach-cone angle $\phi_M$ is given by Mach's
law, $\cos\phi_M= c_s/v_{\rm jet}$. Here,
$c_s$ is the speed of sound and $v_{\rm jet}$ is the jet velocity. If the
leading parton of the jet is light, we have $v_{\rm jet} \simeq 1$, while for
a heavy-flavor leading parton, $v_{\rm jet} < 1$.
Measuring $\phi_M$ and $v_{\rm jet}$, one could in principle extract
$c_s^2 =  dp/ de$ and hence the QGP equation of state (EoS)
$p(e)$, where $p$ is the pressure and $e$ the energy density.

Previous calculations \cite{CasalderreySolana:2006sq,Chaudhuri:2005vc,Renk:2006mv,Betz:2008js,Neufeld,Gubser:2007ga,Noronha:2008un,Betz:2008wy,Betz:2008ka,Li:2009zz} have shown that the formation of 
a double-peak structure on the away-side of soft-hard correlations can be very sensitive to the 
underlying assumptions about the jet-medium interaction \cite{Torrieri:2009mv}. 
In the case of a static medium \cite{CasalderreySolana:2006sq,Betz:2008js,Betz:2008ka}, there 
are two factors that impair the observation of conical correlations, even in a perfect fluid.
First, the thermal smearing at the
freeze-out surface \cite{Cooper:1974mv} broadens the away-side peak
for low-$p_T$ particles \cite{CasalderreySolana:2006sq,Betz:2008js,Noronha:2008un}. 
Second, if the momentum deposited by the jet is larger than a certain threshold,
a diffusion wake moving in the opposite trigger-jet direction may overwhelm any 
signal from the Mach cone \cite{Betz:2008js,Betz:2008ka} and leads to
a single peak on the away-side. Finally, the above discussion neglects
the effects of the expanding medium on
the signal. The strong longitudinal and transverse expansion of the QGP
will distort the Mach-cone signal \cite{Satarov:2005mv}. In Ref.\ \cite{Neufeld:2008eg} 
it was suggested that the diffusion wake may be reduced by transverse flow while longitudinal 
expansion should lead to a broadening of the double-peak structure \cite{Renk:2006mv}. 

In this letter we demonstrate that the double-peak structure 
observed on the away-side of soft-hard correlations 
could be of different origin. In our calculations two effects
conspire to create the observed ``conical'' signal: the averaging
over wakes created by jets in different events and the deflection of
the particles emitted from the wakes by the collective transverse flow.
The resulting ``conical'' signal is found to be
quite robust against variations of the energy-momentum deposition
mechanism and the system size. 
The observed ``cone'' angle is also nearly independent of the jet
velocity, in contrast to the scenario where the double-peak structure
is due to Mach cones.
Therefore, even for events where 
$v_{jet} < c_s$ and proper Mach cones are not formed, the  
intrinsic nonlinearities of the fluid-dynamical equations give rise, after freeze-out, to conical 
structures on the away-side that are strikingly similar to those
produced by a Mach cone.

We use (3+1)-dimensional ideal fluid dynamics  \cite{Rischke:1995pe} to study the evolution of
the QGP,
\begin{equation} \label{1}
\partial_\mu T^{\mu\nu}=S^\nu\,.
\end{equation}
The energy-momentum tensor of the fluid is 
$T^{\mu \nu} = (e+p) u^\mu u^\nu - p\, g^{\mu \nu}$, where $u^\mu$ is
the four-velocity of the fluid. The conservation equations (\ref{1})
are closed by using an EoS, in our case that for an ideal
gas of gluons, $p=e/3$. The source term $S^\nu$ on the right-hand
side of Eq.\ (\ref{1}) is the energy and momentum
deposited by a jet.

We use a transverse initial energy density profile corresponding
to particle creation in central Au+Au and Cu+Cu collisions according to the
Glauber model. The maximum temperature is $T=200$~MeV for Au+Au and $T=176$ MeV for Cu+Cu (since we use an ideal gas 
EoS the exact value of the initial temperature for a given system does not play an important
role in the analysis). In the longitudinal direction, the system is
assumed to be elongated over the whole grid, forming
a cylinder. In this way, we maximize the effect of transverse flow since
there is no additional dilution from longitudinal motion of the fluid.
In our coordinates, the $z$-direction defines the beam axis  
and the associated jet moves along the $x$-direction.

Assuming that the energy and momentum lost by the jet thermalizes
quickly, we use the source term
\begin{eqnarray}
\label{SourceExpandingMedium0}
S^\nu (x) = \int\limits_{\tau_i}^{\tau_f}d\tau 
\frac{dM^\nu}{d\tau} \, \frac{u_\alpha j^\alpha}{u_{0,\beta} j^\beta_0}\,
\delta^{(4)} \left[ x - x_{\rm jet}(\tau) \right],
\end{eqnarray}
with the proper-time interval of the jet evolution $\tau_f - \tau_i$,
the (constant) energy and momentum loss rate $dM^\nu/d\tau=(dE/d\tau,
d\vec{M}/d\tau)$, and the location of the jet $x_{\rm jet}$.
The factor $u_\alpha j^\alpha/u_{0,\beta} j^\beta_0$, where
$j^\alpha$ is the four-current of color charges
($u_0^\beta$, $j_0^\beta$ are the initial four-velocity and
four-current of color charges at the center of
the system), takes into account that the medium expands and cools, thus
reducing the energy-momentum loss rate. In the following, we
shall assume that $j^\alpha \sim T^3 u^\alpha$.
In non-covariant notation, Eq.\ (\ref{SourceExpandingMedium0}) reads
\begin{eqnarray}
\label{SourceExpandingMedium}
S^\nu(t,\vec{x}) &=& \frac{1}{(\sqrt{2\pi}\,\sigma)^3}
\exp\left\{ -\frac{[\vec{x}-\vec{x}_{\rm jet}(t)]^2}{
2\sigma^2}\right\} \nonumber\\
& \times &\left(\frac{dE}{dt},\frac{dM}{dt},0,0\right)
\left[\frac{T(t,\vec{x})}{T_{\rm max}}\right]^3\,.
\end{eqnarray} 
In the following we set $\sigma=0.3$~fm. A temperature cut of
$T_{\rm cut}=130$~MeV is applied 
to ensure that no energy-momentum deposition takes place outside the 
medium. The jet is assumed to move at a constant velocity through the expanding medium [jet deceleration is not expected to lead to significant changes 
after freeze-out \cite{Betz:2008ka}].

In contrast to Ref.\ \cite{Chaudhuri:2005vc}, we assume that 
the parton moving through the QGP stops after it has deposited all of its energy. In the following,
we consider the jet to be generated by a $5$~GeV parton. 
One can compare with experimental data by assuming that, after fragmentation, the
leading hadron carries $\sim 70$\% of the parton's energy,
corresponding to a trigger-$p_T$ of $3.5$~GeV.

Since the experimental analysis can trigger on the jet direction but
not on the space-time location where the jets are formed, 
one has to consider different jet trajectories pointing along the same
direction but originating from different points $(x,y)$ in the
transverse plane \cite{Chaudhuri:2006qk}. We parametrize them as
\begin{eqnarray}
\label{paths}
x &=& r\cos\phi\;,  \hspace*{1cm} y = r\sin\phi\,,
\end{eqnarray}
where $r=5$~fm is chosen to account for the fact that the
trigger jet originates from a point close to the surface. 
A more refined treatment \cite{Renk:2006mv}
will qualitatively lead to the same result. 
We consider different values for the azimuthal angle 
($\Delta\phi=15$~degrees) with
respect to the trigger axis (negative $x$-axis):
$\phi=90, \ldots, 165$~degrees, corresponding to jets travelling 
through the upper half of the transversally expanding medium,
$\phi=180$~degrees, corresponding to a jet travelling along the
(positive) $x$-axis, and $\phi = 195, \ldots, 270$~degrees, corresponding
to jets propagating through the lower half of the medium.

In order to compare to experimental data, one has to convert the
fluid into particles.  In this work, we use the Cooper-Frye (CF)
prescription \cite{Cooper:1974mv} assuming energy-momentum and
entropy conservation across the so-called freeze-out hypersurface
which we take to be a surface at constant time (isochronous freeze-out).
This yields the single-inclusive particle spectrum
$dN/(p_T dp_T dy d\phi)$.

In the experiment, the trajectory of the jet is not
known and one has to measure the azimuthal correlation
between the hard particles produced by the trigger jet and the soft
particles produced by the associated jet traversing the medium. 
In our calculation, we mimic this soft-hard correlation function
by convoluting the single-inclusive particle spectrum
obtained from the CF freeze-out (which only considers the 
away-side particles) with a function representing the near-side jet,
\begin{eqnarray}
f(\phi) &=& \frac{1}{\sqrt{2\pi\Delta\phi^2}}
\exp\left(-\frac{\phi^2}{2\Delta\phi^2}\right)\,,
\end{eqnarray}
(here $\Delta\phi=0.4$), resulting in a 
two-particle correlation function
\begin{eqnarray}
C_2(\phi)&=& A f(\phi) + \int\limits_0^{2\pi}
d\phi^\star \frac{dN}{p_T dp_T dy d(\phi-\phi^\star)} f(\phi^\star)\,,
\end{eqnarray}
where $(A,\Delta\phi)$ are chosen to simulate the  
near-side correlation.  This function is then event-averaged (indicated by
$\langle \cdot \rangle$), background-subtracted, and 
normalized, leading to the averaged two-particle correlation function
\begin{equation} \label{CF}
\langle CF(\phi)\rangle= {\cal N}\,
\left[\langle C_2(\phi) \rangle - 
\frac{d N_{\rm back}}{p_T dp_T dy d\phi}
\right]\,,
\end{equation}
where $d N_{\rm back}/(p_T dp_T dy d\phi)$ is the single-inclusive
particle spectrum for an event without jets and
${\cal N}^{-1} = d N_{\rm back}/(p_T dp_T dy)$.

\begin{figure*}[t]
\centering
  \includegraphics[scale = 0.51]{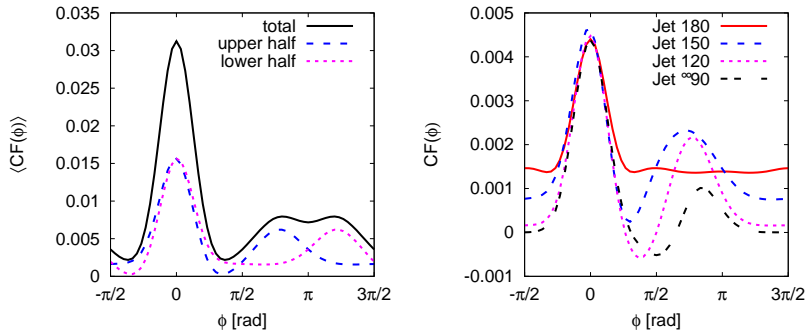}
  \caption{Left panel: The two-particle correlation function (\ref{CF})
   (solid black line) for an associated particle $p_T=2$~GeV.
   The long-dashed blue and short-dashed magenta lines represent the 
   averaged contribution from jets traversing only the upper or the lower 
   half of the medium, respectively. Right panel: the
   unaveraged two-particle correlation function (\ref{CF})
   from four representatively chosen different jet 
   trajectories in the upper half of the medium.}
  \label{FigCF2GeV}
\end{figure*}

Figure \ref{FigCF2GeV} shows the two-particle correlation function 
(\ref{CF}) for $p_T = 2$~GeV.
The jets are assumed to propagate with $v=0.999$, depositing energy-momentum 
into the medium according to Eq.\ (\ref{SourceExpandingMedium}) with
$dE/dt = 1$~GeV/fm and $dM/dt = 1/v\,dE/dt$. 
This case is referred to as ``on-shell'' deposition in the following.
We observe a double-peak structure resembling a Mach-cone signal.
The reason is that the contributions from different jet trajectories,
shown in the right panel of Fig.\ \ref{FigCF2GeV} for jets traversing
the upper half of the medium, add up to a peak at an angle $\phi_w <
180$~degrees (long-dashed blue line in the left panel of Fig.\
\ref{FigCF2GeV}), while the contributions from the 
jets traversing the lower half of the medium 
produce a peak at an angle $\phi_w > 180$~degrees (short-dashed magenta line in 
the left panel of Fig.\ \ref{FigCF2GeV}).
The value of $\phi_w$ depends on how much the transversally expanding
medium can deflect the matter in the disturbance caused by the jet. 
This, in turn, depends on the local flow velocity (both its magnitude
and direction with respect to the associated jet) and temperature.
The gap between the two peaks on the 
away-side depends on the $p_T$ of associated particles. For small $p_T$, 
thermal smearing is large and the two peaks merge into a single
broad away-side peak \cite{Betz:2009su}.
The same flow induced peak-to-valley ratio is found within
0.3\% if a coarser $\Delta\phi=30$~degrees initial jet sampling is used. 
The robustness of the numerical results in Fig.\ \ref{FigCF2GeV} to 
numerical details underlines the universality of the radial flow mechanism
as the primary source of non-Mach away-side azimuthal correlations.

\begin{figure*}[t]
\centering
 \includegraphics[scale = 0.51]{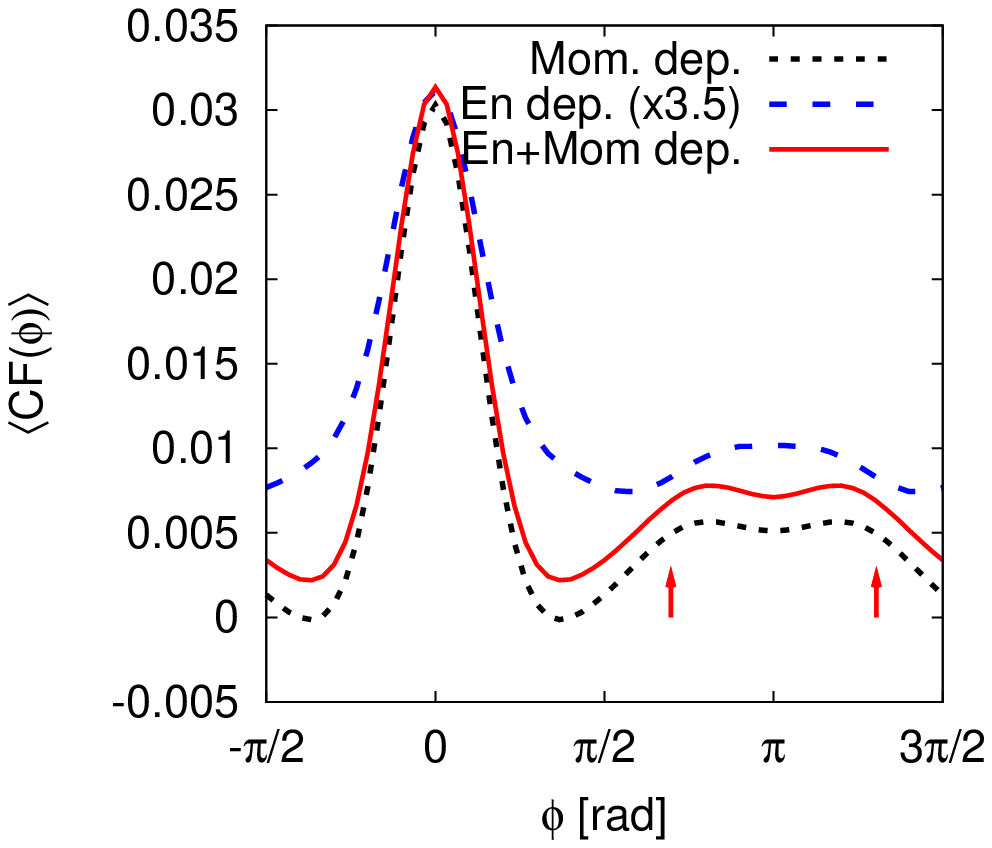}
 \includegraphics[scale = 0.51]{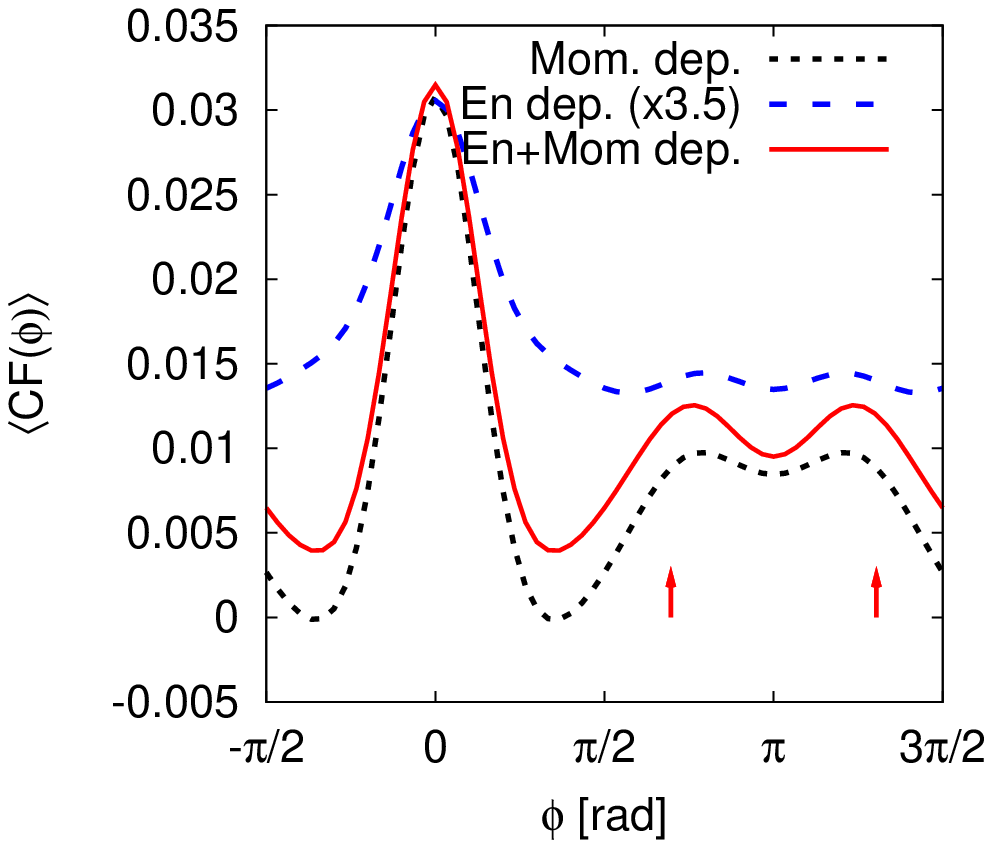}
  \caption{The two-particle correlation function (\ref{CF}) for jets
propagating with $v=0.999$. 
Solid red line: on-shell energy-momentum deposition.
Short-dashed black line: pure momentum deposition.
Long-dashed blue line: pure energy deposition (scaled by a factor
3.5).
Left panel:  $p_T=2$~GeV. Right panel: $p_T=3$~GeV. 
Arrows indicate the emission angle obtained by Mach's law.}
  \label{FigAllDep}
\end{figure*}

Figure \ref{FigAllDep} shows that a similar double-peak structure as
in the on-shell deposition scenario can also be obtained for a pure momentum
deposition scenario with $dE/dt = 0$, and $dM/dt = 1.0/v$~GeV/fm. 
Both scenarios lead to approximately the same apparent ``cone'' angle. On the
other hand, due to thermal smearing a pure energy deposition scenario (with $dE/dt = 1$~GeV/fm
and $dM/dt=0$) 
only exhibits a single broad peak for $p_T = 2$~GeV. For 
a larger $p_T = 3$~GeV, the double-peak structure reappears, with a similar
``cone'' angle as in the other deposition scenarios. 
Thus, in an expanding medium, the apparent ``cone'' shape resulting
from averaging over many events is universal
to all energy-momentum deposition scenarios, in contrast to the case
of a static background \cite{Torrieri:2009mv}.
Therefore, we cannot draw
conclusions about the jet deposition mechanism, although the pure energy loss scenario seems to be
disfavored by the data, since the two peaks appear only at a higher $p_T$.

\begin{figure}[h]
\centering
  \includegraphics[scale = 0.51]{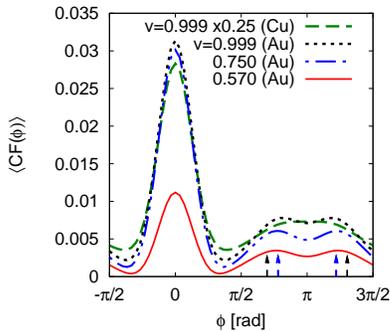}
  \caption{The two-particle correlation function (\ref{CF}) for jets
travelling at $v=0.999$ in central Cu+Cu collisions (long-dashed green line) and
central Au+Au collisions (short-dashed black line), $v=0.75$ (dash-dotted blue line), 
  and $v=0.57$ (solid red line) for  $p_T=2$~GeV. For the supersonic jets,
  the arrows indicate the emission angles obtained by Mach's law.}
  \label{FigAllVelocities}
\end{figure}

Finally, we demonstrate that the conical emission angle observed in
Figs.\ \ref{FigCF2GeV} and \ref{FigAllDep} also appears
for subsonic jets (which should not be the case if it was due to a
true Mach cone). We consider a bottom quark with mass
$M=4.5$~GeV, propagating with $v=0.57< c_s$ through the medium.
The results for the correlation function (\ref{CF}), computed with
the on-shell energy-momentum deposition scenario with
$dE/dt=1$~GeV, are shown in Fig.\ \ref{FigAllVelocities},
where, for comparison, we also show results for $v=0.75$.

The same double-peak emission structures are found for
all jet velocities, even for the subsonic jet, with nearly equal apparent
``cone'' angles.
A comparison between the away-side yield in Au+Au and Cu+Cu is also shown
in Fig.\ \ref{FigAllVelocities}. Remarkably, both systems 
are predicted in this scenario to display a very similar away-side
shoulder width in qualitative agreement with recent experimental
results from the PHENIX collaboration \cite{Adare:2006nr}. 
While the absolute normalization of $\langle CF(\phi) \rangle$ is sensitive
to the freeze-out temperature $T_f$, we found that the double-peak
structures in Au+Au are insensitive to variations of $T_f$ between 130~MeV and
160~MeV. For Cu+Cu the weak dip at $\phi=\pi$ is filled for $p_T=2$GeV
but reappears at higher $p_T$ while the shoulder width remains similar to the Au+Au case.
Three-particle correlations will be
presented elsewhere. 

In conclusion, we have shown that a double-peak structure on the away side of
soft-hard correlations can originate generally from the coupling of jet
fragments to the background transverse collective flow. In addition,
the apparent width of the away-side shoulder correlation
is ``universal'' in the sense
that it is insensitive to the details of the energy-momentum
deposition mechanism as well as to the system size, and that is similar for both supersonic {\em and subsonic}
``jets''. This prediction can be readily tested experimentally by comparing
soft-hard correlations induced by heavy-flavor tagged jets \cite{Antinori:2005tu} with those induced
by light-flavor jets at RHIC and LHC.   
  
B.B.\ is supported by the Alexander von Humboldt foundation via a Feodor Lynen fellowship. 
M.G., J.N., and B.B.\ acknowledge support from DOE under
Grant No.\ DE-FG02-93ER40764. G.T.\ acknowledges support from the Helmholtz International 
Center for FAIR within the framework of the LOEWE program.

\end{document}